\documentclass[cameraready]{Interspeech}
\usepackage[subrefformat=parens]{subcaption}
\usepackage{multicol}
\usepackage{multirow}
\usepackage{cite}
\title{Scalable Direction-Following TTS \\via Voice Impression-Guided Pseudo Triplet Construction}

\author[affiliation={}]{Kenichi}{Fujita}
\author[affiliation={}]{Yusuke}{Ijima}

\address{NTT, Inc., Japan}

\email{kenichi.fujita@ntt.com}


\usepackage{comment}

\begin{document}

\setlength\textfloatsep{7pt} 
\setlength\dbltextfloatsep{7pt} 
\setlength\floatsep{7pt} 
\setlength\abovecaptionskip{2pt} 
\setlength\belowcaptionskip{2pt} 
\captionsetup[subfloat]{aboveskip=2pt,belowskip=4pt} 

\maketitle

\begin{abstract}
Voice actors often re-read the same script while modifying their delivery in response to performance directions.
We study this setting as direction-following TTS, where a system generates a new utterance that reflects a given direction relative to a reference utterance while preserving speaker identity and linguistic content.
A key challenge is the lack of training data capturing such relative modifications.
To address this, we propose a scalable pseudo-triplet construction pipeline that generates~(reference utterance, direction text, modified utterance) triplets.
It generates controlled style variations using an impression-controllable TTS model and uses an LLM to produce natural language directions from estimated impression differences.
Experimental results demonstrate that pseudo-triplets alone enable stable speaker-preserving modification, and that combining pseudo and recorded data further improves direction alignment while maintaining speaker similarity.

\end{abstract}

\section{Introduction}
Beyond linguistic content, speech conveys para- and non-linguistic information such as speaking style and perceived voice impressions~\cite{SCHULLER20134,cowen2019mapping}.
In acting, performers adjust their delivery in response to performance directions from a director, re-reading the same script with different speaking styles.
Directors often provide feedback on the line, guiding voice actors to modify their speaking style relative to their previous take~\cite{Schmidt23_showing}.

This scenario naturally motivates a text-to-speech synthesis~(TTS) setting that models direction-driven re-reading of a fixed script~\cite{kanagawa25_ssw}.
In this setting, a user provides a script, a reference utterance, and a natural language direction describing how the performance should change.
The system generates a new utterance that preserves speaker identity and content while reflecting the stylistic modification.
We refer to this task as \textit{direction-following TTS}. 
Unlike conventional style-conditioned TTS, this task models a direction-conditioned transformation of a reference utterance, requiring paired pre- and post-modification utterances with corresponding direction text.
The reference utterance and its direction-modified counterpart are termed as the pre-modification utterance (\textit{pre-mod utterance}) and the post-modification utterance (\textit{post-mod utterance}), respectively.

A central challenge is the lack of training triplets consisting of a \textit{pre-mod utterance}, a direction specifying the modification, and a corresponding \textit{post-mod utterance} for the same script.
Large-scale speech corpora used to train zero-shot TTS models~\cite{ma24d_interspeech,langman25_interspeech} typically contain only a single reading of each script and thus do not capture relative stylistic changes.
Even when corpora with multiple recordings of the same script are available~\cite{zhou2021emotional,yamagishi2005HMM}, they are typically small and lack directions describing the intended modification.
This scarcity hinders not only data-intensive generative models such as diffusion- and flow-based approaches~\cite{NEURIPS2023_voicebox,chen-etal-2025-f5}, but also robust speaker-preserving style transformation across diverse speakers.

\begin{figure}[tb]
  \centering
  \includegraphics[width=1.0\linewidth]{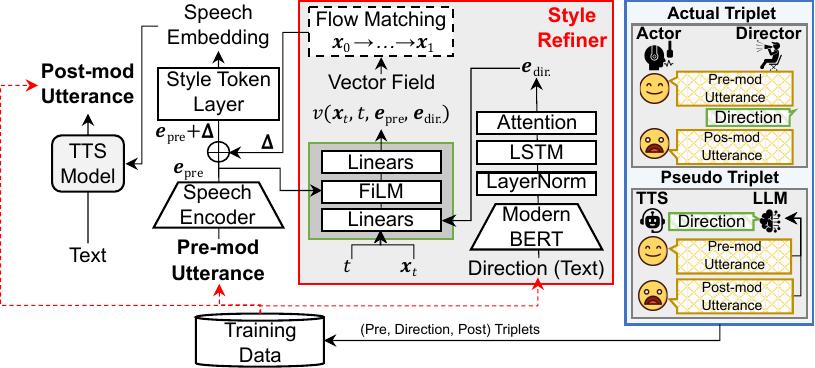}
  \caption{Overview of refinement with direction and actual and pseudo training data.}
  \label{fig:overview}
\end{figure}

To address this issue, we construct pseudo training triplets.
An impression-controllable TTS model~\cite{fujita25_interspeech,Ohmura25_libri} serves as a proxy performer to generate paired \textit{pre-mod} and \textit{post-mod utterances}.
A large language model (LLM) produces natural language directions from estimated impression differences, forming pseudo triplets that capture direction-induced variation.
Our approach learns a transformation from \textit{pre-mod} to \textit{post-mod} utterances (Fig.~\ref{fig:overview}).
This differs from conventional text-prompted TTS and speech editing systems~\cite{ICLR2024_fed1ea8d,huang2024instructspeech, ji-etal-2025-controlspeech,Jin24_speechcraft,Yong26_ov,Yun26_isse,Yao24_promptvc,chen2026flexivoiceenablingflexiblestyle}, which focus on absolute style conditioning or prompt-based control.
We demonstrate the effectiveness of this framework in a zero-shot TTS through objective and subjective evaluations, including an exploratory LLM-assisted evaluation framework. 
Audio examples are available on our demo page\footnote{\url{https://ntt-hilab-gensp.github.io/IS2026pseudo/}\label{foot:sample_page}}.

\begin{figure*}[tb]
  \centering
  \includegraphics[width=0.9\linewidth]{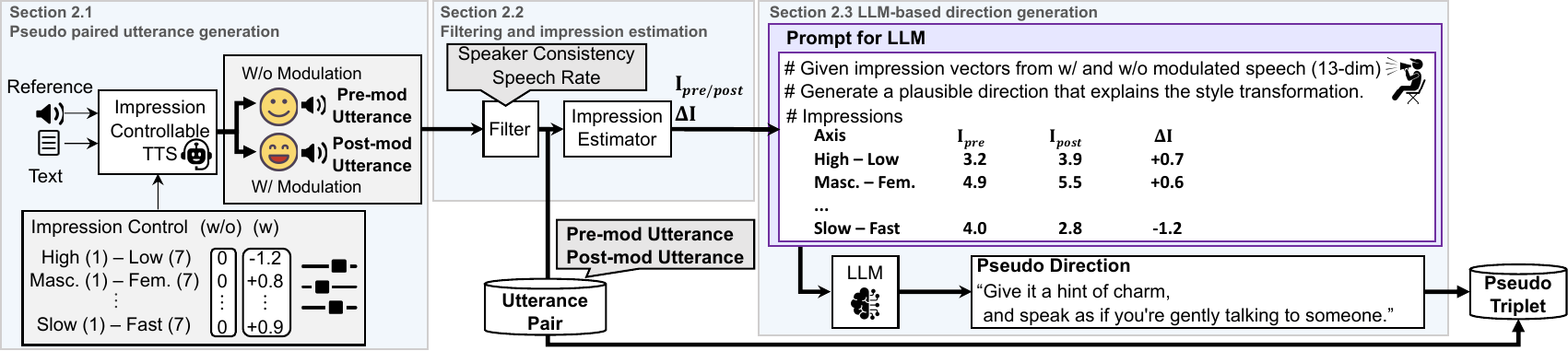}
  \caption{Overview of pseudo data generation.}
  \label{fig:pseudo_gen}
\end{figure*}

\section{Data design for direction-following}
We use two complementary data sources: pseudo data generated by TTS and recorded data from interactive recordings. 
Pseudo data enables scalability, while recorded data captures authentic human responses.
This section describes pseudo-triplet construction (Fig.~\ref{fig:pseudo_gen}); recorded data are described in Sect.~\ref{sec:expset_actualdata}.

\subsection{Pseudo paired utterance generation (Fig.~\ref{fig:pseudo_gen} left)}\label{sec:pseudo_pair_generation}
Pseudo paired speech generation requires fine-grained style control while preserving speaker identity.
We therefore use an impression-controllable zero-shot TTS method~\cite{fujita25_interspeech,Ohmura25_libri}.
For each script and reference utterance, we synthesize a baseline utterance and a style-modified counterpart, treated as the \textit{pre-mod} and \textit{post-mod utterances}, respectively.

Impression is controlled via a low-dimensional impression vector following prior work~\cite{fujita25_interspeech,Ohmura25_libri}.
The vector comprises 13 continuous dimensions representing the intensity of antonym-based voice impressions rated on a subjective 7-point Likert scale.
Eleven dimensions follow validated impression axes~\cite{fujita25_interspeech,kasuya1999_extraction,Mizuki_Nagano2024e24.14}: high--low pitched, masculine--feminine, clear--hoarse, calm--restless, powerful--weak, youthful--elderly, thick--thin, tense--relaxed, dark--bright, cold--warm, and slow--fast.
We extend this representation with two axes, fluent--hesitant and emotional--neutral, to better reflect performance directions.

\subsection{Filtering and impression estimation (Fig.~\ref{fig:pseudo_gen} center)}
We filter out pairs with excessive speaker drift or abnormal speaking-rate differences.
Speaker consistency is measured by cosine similarity between speaker embeddings and speaking-rate differences.
Pairs with near-identical speaker embeddings are also excluded as negligible variation provides little guidance for modeling direction-dependent modification.

For the remaining pairs, we estimate a 13-dimensional impression vector for each utterance using an impression estimator.
The estimator predicts the same antonym-based dimensions defined above.
Let $\mathbf{I}_{pre}$ and $\mathbf{I}_{post}$ denote the impression vectors of the \textit{pre-mod} and \textit{post-mod utterances}, respectively.
The impression difference is $\Delta \mathbf{I} = \mathbf{I}_{post} - \mathbf{I}_{pre}$.

\subsection{LLM-based direction generation (Fig.~\ref{fig:pseudo_gen} right)}
Given the estimated impression vectors $\mathbf{I}_{pre}$ and $\mathbf{I}_{post}$ and their difference $\Delta \mathbf{I}$, we generate a natural language performance direction using a LLM.
Unlike conventional absolute labeling of a single utterance~\cite{ji-etal-2025-controlspeech,chen2026flexivoiceenablingflexiblestyle}, our direction generation is conditioned on the relative difference between two utterances, enabling the LLM to describe the transformation from \textit{pre-mod} to \textit{post-mod}.
The LLM is prompted to act as a voice director and infer a plausible direction that explains the impression change. 
Although guided by $\Delta \mathbf{I}$, the generated directions are not restricted to explicit axis-level descriptions but may include compositional or context-dependent acting instructions (e.g., ``maintaining a restrained tone with subtle hesitation''). 
Thus, impression vectors serve only as intermediate cues rather than a target style representation. 
The resulting triplet (\textit{pre-mod utterance}, pseudo direction, \textit{post-mod utterance}) forms a pseudo training sample.
A prompt example is available on our demo page\footref{foot:sample_page}.

\section{Direction-conditioned style refiner}
We introduce a direction-conditioned style refiner that modifies the speech embedding (Fig.~\ref{fig:overview}).
The refiner operates in the speech embedding space, while the backbone TTS model remains fixed.
This allows it to focus on direction-dependent style refinement.
At inference, the predicted embedding modification is added to an embedding from the \textit{pre-mod utterance}.

Let $e_{\text{pre}}$ and $e_{\text{post}}$ denote embeddings extracted from the \textit{pre-mod} and \textit{post-mod utterance}, respectively.
We define the target modification as $\Delta = e_{\text{post}} - e_{\text{pre}}$, treating style refinement as a local additive approximation in the speech embedding space without assuming global linearity.
In \textit{direction-following TTS}, a given direction may correspond to multiple plausible embedding shifts depending on speaker characteristics.
Deterministic regression would collapse such variability into a conditional mean, leading to conservative updates.
We therefore adopt rectified flow matching~\cite{liu2022flow} to model a direction-conditioned stochastic vector field in the embedding space.
A noise vector $x_0$ is sampled from a Gaussian distribution, and an intermediate state is constructed as
\[
x_t = (1 - t) x_0 + t \Delta, \quad t \in [0, 1].
\]
The refiner predicts the constant velocity field $v_\theta(x_t, t, e_{\rm{pre}},e_{\rm{dir}})$, conditioned on the \textit{pre-mod utterance} $e_{\rm{pre}}$, direction representation $e_{\rm{dir}}$ and time $t$, and is trained to match the target velocity $\Delta - x_0$.
Training minimizes a flow matching objective, with auxiliary losses encouraging directional consistency and magnitude alignment between the predicted and target modifications.

\begin{figure*}[t]
  \centering
  \subfloat[Results for seen speakers]{%
    \includegraphics[width=0.47\linewidth]{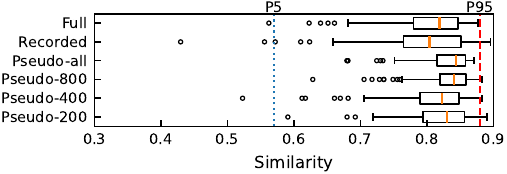}
  }\hfill
  \subfloat[Results for unseen speakers]{%
    \includegraphics[width=0.47\linewidth]{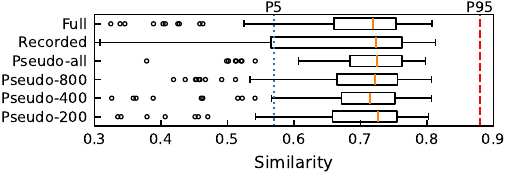}
  }
  \vspace{-2mm}
  \caption{Speaker similarity to pre-mod utterances using ECAPA-TDNN embeddings. P5/P95 show reference stats from recorded pairs.}
  \label{fig:obj_similarity}
  \vspace{-2mm}
\end{figure*}

\section{Experimental setup}
\subsection{Datasets}
\subsubsection{Pseudo data}
For pseudo data generation, we used in-house Japanese speech data from 1,600 speakers.
Ten recorded utterances were randomly selected per speaker, and for each selected utterance, ten paired utterances with different sentences were generated using an impression-controllable TTS model with and without impression control.
Impression control was applied by randomly selecting three impression dimensions whose pairwise correlation coefficients (estimated on the impression estimator's outputs) were below 0.7 and sampling control values in the range of \textminus2 to +2 on the original 7-point scale.
This resulted in 160,000 initially generated utterance pairs before filtering.

Pairs were filtered on the basis of speaker consistency~(ECAPA-TDNN cosine similarity~\cite{speechbrain}: 0.80--0.95) and speaking rate ratios~(0.85--1.15) to ensure stable speaker identity while retaining moderate style variation.
After filtering, we obtained 74,619 utterance pairs.

For each utterance pair, up to five natural language performance directions were generated to increase diversity per pair using Qwen3-Next-80B-A3B-Instruct\footnote{\url{https://huggingface.co/Qwen/Qwen3-Next-80B-A3B-Instruct}\label{foot:qwen}}.
Malformed directions (e.g., overly long text or abnormal symbols) were removed via simple text-filtering.
Each valid direction combined with its corresponding speech pair formed a pseudo triplet (\textit{pre-mod utterance}, direction, \textit{post-mod utterance}).

After text filtering, the final dataset consisted of 350,617 pseudo triplets corresponding to 127.6 hours of unique speech.
The data were split into 346,488 triplets for training and 4,129 for validation using 30 held-out speakers.
To investigate the effect of the pseudo-data scale, we constructed speaker-diversity subsets with 200, 400, and 800 speakers by sampling from the full 1,600-speaker pseudo dataset, comprising 43,131, 86,643, and 175,091 triplets, respectively.

\subsubsection{Actual recording data}\label{sec:expset_actualdata}
In addition to pseudo data, we collected a small dataset through interactive recording sessions with two professional voice actors (one male and one female), resulting in 8.9 hours of speech~(6,899 utterance pairs).
Performance directions were prepared in advance using a LLM following the protocol of Kanagawa et al.~\cite{kanagawa25_ssw} to ensure consistency and scalability.
During recording, the actor re-read the same script while modifying their performance according to the provided direction.

To increase linguistic variability, direction text augmentation was applied using the Easy, Medium, and Hard prompt templates proposed by Maini et al.~\cite{maini2024rephrasing}, excluding the Q/A style.
Augmented directions were generated using LLM\footref{foot:qwen}, with templates translated into Japanese.
Although small, this dataset captures natural human responses to performance directions that are difficult to reproduce through synthesis alone.

\subsection{Supporting models and components}
Our framework has three main components: a voice impression estimator, an impression-controllable TTS model, and
a direction-conditioned style refiner.
The impression estimator is shared across pseudo data construction and impression-controllable TTS training to ensure consistent impressions.

The voice impression estimator maps speech to the 13-dimensional impression representation described in Sect.~\ref{sec:pseudo_pair_generation} and is trained on human subjective ratings.
It generalizes to unseen speakers, achieving a root mean squared error of 0.40 on held-out utterances (5,481 utterances from 49 speakers), and is fixed for all stages of pseudo data construction, TTS training, and evaluation.
Architecture and training details follow~\cite{fujita25_interspeech,Ohmura25_libri}.

The impression-controllable TTS model is based on a FastSpeech2~\cite{ren2020fastspeech} backbone with a frozen HuBERT-based jointly trained speech encoder~\cite{hsu2021hubert,fujita2023zeroshot}, trained under a GAN~\cite{kanagawa19_ssw,peng2018variational}.
It was trained on 27,000 hours of multi-speaker in-house Japanese speech, followed by training an impression control module using impression vectors predicted by the impression estimator~\cite{fujita25_interspeech,Ohmura25_libri}.
Waveform generator is HiFi-GAN (V1)~\cite{NEURIPS2020_c5d73680}. 

The style refiner is trained on top of the fixed backbone TTS model.
For direction encoding, we use ModernBERT-Ja-310M~\cite{modernbert-ja}.
The refiner is optimized with Adam~\cite{kingma-Adam} (learning rate 0.01, batch size 32) for up to 1M steps.
To analyze the effects of data composition and speaker diversity, we train separate models using pseudo data only~(\textit{Pseudo-all}), actual recording data only~(\textit{Recorded}), their combination~(\textit{Full}), and pseudo data subsets with 200, 400, and 800 speakers~(\textit{Pseudo-200/400/800}).
For each configuration, the model with the lowest validation loss is selected.

\section{Results}
\subsection{Objective evaluation}\label{sec:obeval}
We conducted objective evaluations of naturalness, speaker similarity, and direction alignment.
We evaluated four speakers: two seen speakers from actual recording dataset and two unseen speakers not used to train the supporting models (refiner, impression estimator, and impression-controllable TTS), selected from the HiFi-CAPTAIN~\cite{hi-fi-captain} test set.
Each group had one male and one female speaker.

For objective evaluation, we generated 15,000 direction-conditioned utterances per speaker.
To isolate the effect of performance directions, \textit{pre-mod utterances} were synthesized using the fixed backbone TTS model.
For each speaker, 100 recorded utterances were selected as sources.
For each source, three sentences were randomly selected from a fixed pool of ten sentences and synthesized as \textit{pre-mod utterances}, resulting in 300 utterances per speaker.
We then generated a fixed set of 50 directions using GPT-5.2~\cite{gpt5_2}, prompting it to act as a voice director.
The same 50 directions were applied to all speakers, and each \textit{pre-mod utterance} was combined with all directions.
To avoid potential bias from pseudo-data construction, we used a different LLM that used for pseudo direction generation.

We first evaluated speaker similarity using the cosine similarity of ECAPA-TDNN-based speaker embeddings between \textit{pre-mod} and \textit{post-mod utterances}.
Reference statistics from actual recorded utterance pairs yielded 5th and 95th percentiles of 0.57 and 0.89, representing the typical range for intra-speaker variation.
Figure~\ref{fig:obj_similarity} shows the results summarized per direction.
For each speaker, similarity scores were averaged over the 300 reference utterances per direction, and the resulting 50 direction-level means were used to construct the box plots, mitigating utterance-level variability.
For seen speakers, similarity scores were generally stable across conditions, with fewer instances falling below the 5th percentile threshold.
Similar trends were observed for unseen speakers.
For unseen speakers, the \textit{Recorded} condition showed large variance, with similarity scores often falling below 0.57, indicating unstable identity preservation.
In contrast, \textit{Pseudo} conditions were more robust, and increasing speaker diversity reduced extreme speaker drift.
The \textit{Full} condition remained more stable than \textit{Recorded}, balancing robustness and variability.

Next, we evaluated speech naturalness using UTMOSv2~\cite{baba2024utmosv2}~(Table~\ref{table:obj_llm}).
Source recorded utterances scored $2.67\pm0.01$~(seen) and $2.97\pm0.01$ (unseen).
Across all configurations, direction refinement did not noticeably degrade naturalness, with scores comparable to the recorded utterances.

We finally evaluated direction alignment automatically using an LLM\footref{foot:qwen}.
While prior work on absolute style conditioning directly evaluates text–speech alignment~\cite{Yong26_ov}, evaluating direction-following between two utterances is more challenging~\cite{fujita26_rie}. 
We therefore approximated alignment by comparing estimated impression changes with the semantic content of the direction.
For each \textit{pre-} and \textit{post-mod utterance} pair,  impression vectors were estimated and provided to the LLM with the direction text.
The LLM then rated alignment on a 1–5 scale (1: Not aligned at all to 5: Perfectly aligned).
A prompt example is available on our demo page.\footref{foot:sample_page}
Since the same impression estimator is used in pseudo-data construction, this metric should be interpreted as a diagnostic proxy.
To verify its validity, we conducted two sanity checks.
Actual recorded direction-speech pairs yielded an alignment score of $3.12 \pm 0.03$ (95\% CI).
Replacing directions with semantically opposite or unrelated ones generated by a separate LLM\footnote{\url{https://huggingface.co/Qwen/Qwen3-30B-A3B-Instruct-2507}}
reduced scores to $1.87 \pm 0.02$ and $1.91 \pm 0.02$, confirming sensitivity to mismatches.

As shown in Table~\ref{table:obj_llm}, \textit{Full} and \textit{Recorded} conditions achieved higher alignment, while \textit{Pseudo} conditions exhibited more conservative modulation.
Notably, \textit{Full} attained alignment comparable to \textit{Recorded} while maintaining improved speaker robustness, indicating complementary benefits of pseudo and recorded data.
Increasing pseudo-data scale further stabilized speaker identity preservation without substantially degrading alignment.

\begin{table}[tb]
    \caption{Results of objective evaluations with 95\% confidence interval. Bold shows best scores. \textit{Pre-mod} represents UTMOS of \textit{Pre-mod utterances}. }
    \label{table:obj_llm}
    \centering
    \setlength{\tabcolsep}{4pt}
    \resizebox{\linewidth}{!}{%
        \begin{tabular}{@{}l@{\hspace{3pt}}c*{4}{c}@{}}
            \toprule
            & \multicolumn{2}{c}{Seen} & \multicolumn{2}{c}{Unseen}\\
            \cmidrule(l{\tabcolsep}){2-3}\cmidrule(l{\tabcolsep}){4-5}
            Conditions & UTMOS & LLM & UTMOS & LLM \\
            \midrule
            \textit{Pre-mod}        & $2.95\pm0.01$ & \textendash & $2.97\pm0.01$ & \textendash   \\
            Full        & $ 2.96\pm0.01$ & $\bf 3.79\pm0.01$& $2.97\pm0.01$ & $3.24\pm0.02$   \\
            Recorded    & $2.94\pm0.01$ & $3.78\pm0.01$& $2.95\pm0.01$ & $\bf 3.37\pm0.02$  \\
            Pseudo-all  & $2.95\pm0.01$ & $3.67\pm0.01$& $2.99\pm0.01$ & $3.12\pm0.02$    \\
            Pseudo-800  & $2.94\pm0.01$ & $3.64\pm0.01$& $ 3.00\pm0.01$ & $3.15\pm0.02$   \\
            Pseudo-400  & $ 2.96\pm0.01$ & $3.66\pm0.01$& $2.99\pm0.01$ & $3.19\pm0.02$  \\
            Pseudo-200  & $ 2.96\pm0.01$ & $3.67\pm0.01$& $ 3.00\pm0.01$ & $3.17\pm0.02$  \\
            \bottomrule
        \end{tabular}
    }
\end{table}

\begin{table}[tb]
    \caption{Results of subjective evaluation with 95\% confidence interval. Bold shows best scores. }
    \label{table:subjective}
    \centering
    \setlength{\tabcolsep}{4pt}
    \resizebox{\linewidth}{!}{%
        \begin{tabular}{@{}l@{\hspace{3pt}}c*{4}{c}@{}}
            \toprule
            & \multicolumn{2}{c}{Seen} & \multicolumn{2}{c}{Unseen}\\
            \cmidrule(l{\tabcolsep}){2-3}\cmidrule(l{\tabcolsep}){4-5}
            Conditions & SMOS & AlignMOS & SMOS & AlignMOS \\
            \midrule
            Full        & $3.35\pm0.08$ & $3.22\pm0.07$& $3.22\pm0.07$ & $3.32\pm0.07$   \\
            Recorded    & $2.67\pm0.08$ & $\bf 3.50\pm0.07$& $2.63\pm0.08$ & $\bf 3.48\pm0.07$  \\
            Pseudo-all  & $\bf 3.54\pm0.08$ & $3.08\pm0.07$& $\bf 3.24\pm0.07$ & $3.18\pm0.07$    \\
            \bottomrule
        \end{tabular}
    }
\end{table}

\subsection{Subjective evaluation}
We conducted subjective evaluations of speaker similarity and direction alignment.
The results are summarized in Table~\ref{table:subjective} as SMOS and AlignMOS.
We compared three conditions: \textit{Full}, \textit{Recorded}, and \textit{Pseudo-all}.
The smaller pseudo-data variants (Pseudo-200/400/800) were excluded to limit evaluation burden, as objective results showed consistent trends across scales.

For each speaker, we evaluated 60 triplets.
These were selected from the objective evaluation set by choosing three \textit{pre-mod utterances} and 20 representative directions, resulting in 3 × 20 combinations per speaker.
The evaluation was conducted via crowdsourcing with 258 and 208 participants, respectively, and each sample was rated by at least ten listeners.

To evaluate speaker similarity, participants rated whether  \textit{pre-mod} and \textit{post-mod utterances}
were spoken by the same speaker using a 5-point Likert scale
(1: not the same speaker, 5: definitely the same speaker; SMOS).
The results in Table~\ref{table:subjective} are consistent with the objective evaluation for both seen and unseen speakers.
\textit{Recorded} exhibited lower speaker similarity, indicating unstable speaker identity preservation.
In contrast, \textit{Full} and \textit{Pseudo-all} showed higher scores.
Although \textit{Pseudo-all} slightly outperformed \textit{Full}, this likely reflects its more conservative style modulation, resulting in smaller changes between \textit{pre-mod} and \textit{post-mod utterances} and thus higher similarity.

For direction alignment, participants listened to pairs of \textit{pre-mod} and \textit{post-mod utterances} and rated how the change matched the given direction on a 5-point Likert scale (1: not aligned at all, 5: perfectly aligned; AlignMOS).
Participants were also instructed to assign a score of 1 if speaker identity changed significantly ensuring alignment was judged under stable identity.
From Table~\ref{table:subjective}, \textit{Recorded} achieved the highest alignment scores, followed by \textit{Full} and \textit{Pseudo-all}.
Its higher scores likely reflect larger style variations, making changes more perceptible despite reduced speaker similarity.
In contrast, \textit{Pseudo-all} exhibited slightly lower alignment, consistent with more conservative style modulation.
\textit{Full} lies between the two, suggesting that combining pseudo and recorded data balances expressive variation and identity stability.

To better understand the relatively conservative modulation observed in \textit{Pseudo-all}, we analyzed the training data by measuring the absolute change in F0 between \textit{pre-mod} and \textit{post-mod utterances}. 
Utterance pairs from professional recordings (\textit{Recorded}) showed larger absolute changes in mean ln F0 ($0.14 \pm 0.15$, mean $\pm$ SD) than pseudo-generated pairs (\textit{Pseudo}, $0.05 \pm 0.08$). 
A similar trend was observed for pitch variability: the absolute change in the standard deviation of ln F0 was also larger for \textit{Recorded} ($0.057 \pm 0.063$) than for \textit{Pseudo} ($0.024 \pm 0.041$).
These results suggest that pseudo data exhibits smaller prosodic variation than professionally recorded performances, which may partly explain the more conservative style modulation in \textit{Pseudo-all}. 
This difference likely reflects the broader expressive range of professional acting speech while still preserving speaker identity, rather than a limitation of the pseudo-triplet approach. Similar tendencies may arise in other TTS- and VC-based~\cite{Yun26_isse} pseudo data generation frameworks.

Overall, \textit{Full} provides the most favorable trade-off between direction alignment and speaker identity preservation.
Notably, \textit{Pseudo-all} alone already enables stable identity-preserving refinement with reasonable direction alignment, demonstrating the effectiveness of the proposed pseudo-triplet construction.

The subjective alignment results are broadly consistent with the LLM-based evaluation in Sect.~\ref{sec:obeval}, lending support to the validity of the proposed objective metric. 
This agreement also provides indirect validation of the impression-difference representation as a meaningful basis for approximating direction speech alignment, consistent with recent work using LLMs as evaluators in other modalities~\cite{chen2024mllmasajudge,li-etal-2025-generation}.

\section{Conclusion}
We proposed \textit{direction-following TTS} to model relative performance modification. 
To address the scarcity of paired direction data, we constructed pseudo triplets via impression-controlled synthesis and LLM-based direction generation.
Objective and subjective evaluations showed that pseudo data improves speaker robustness, recorded data enhances direction alignment, and their combination achieves the best balance between stability and expressiveness. 
Future work will investigate how to integrate human expressiveness with pseudo data.

\clearpage
\section{Disclosure of generative AI use}
During the preparation of this manuscript, the authors used ChatGPT (OpenAI) for language editing and refinement. 
All outputs were reviewed and revised as necessary by the authors, who take full responsibility for the content of this publication.
\bibliographystyle{IEEEtran}
\bibliography{refs}

@inproceedings{ren2020fastspeech,
  title={{FastSpeech} 2: Fast and High-Quality End-to-End Text to Speech},
  author={Ren, Yi and Hu, Chenxu and Tan, Xu and Qin, Tao and Zhao, Sheng and Zhao, Zhou and Liu, Tie-Yan},
  booktitle={Proc. ICLR},
  year={2020}
}

@article{hsu2021hubert,
  title={{HuBERT}: Self-supervised speech representation learning by masked prediction of hidden units},
  author={Hsu, Wei-Ning and Bolte, Benjamin and Tsai, Yao-Hung Hubert and Lakhotia, Kushal and Salakhutdinov, Ruslan and Mohamed, Abdelrahman},
  journal={TASLP},
  volume={29},
  pages={3451--3460},
  year={2021},
  publisher={IEEE}
}

@inproceedings{NEURIPS2020_c5d73680,
 author = {Kong, Jungil and Kim, Jaehyeon and Bae, Jaekyoung},
 booktitle = {Proc. NeurIPS},
 title = {{HiFi-GAN}: Generative Adversarial Networks for Efficient and High Fidelity Speech Synthesis},
 year = {2020},
 pages = {17022--17033}, 
 volume = {33}
}

@INPROCEEDINGS{fujita2023zeroshot,
  author={Fujita, Kenichi and Ashihara, Takanori and Kanagawa, Hiroki and Moriya, Takafumi and Ijima, Yusuke},
  booktitle={Proc. ICASSP Workshops (ICASSPW)}, 
  title={Zero-Shot Text-to-Speech Synthesis Conditioned Using Self-Supervised Speech Representation Model},  
  year={2023},
}

@InProceedings{kingma-Adam,
  title = 	 {Adam: A method for stochastic optimization},
  author =       {Kingma, Diederik P and Ba, Jimmy},
  booktitle = 	 {Proc. ICLR},
  year = 	 {2015},
}

@article{Mizuki_Nagano2024e24.14,
  title={The influence of semantic primitives in an emotion-mediated willingness to buy model from advertising speech},
  author={Mizuki Nagano and Yusuke Ijima and Sadao Hiroya},
  journal={Acoustical Science and Technology},
  volume={46},
  number={1},
  pages={87-95},
  year={2025},
}

@ARTICLE{kasuya1999_extraction,
  author={Hiroshi Kido and Hideki Kasuya},
  journal={Journal of ASJ}, 
  title={Extraction of everyday expression associated with voice quality of normal utterance }, 
  year={1999},
  volume={55},
  number={6},
  pages={405-411},
  note={(In Japanese with English title)}
}

@inproceedings{kanagawa19_ssw,
  title     = {Multi-Speaker Modeling for {DNN}-based Speech Synthesis Incorporating Generative Adversarial Networks},
  author    = {Hiroki Kanagawa and Yusuke Ijima},
  year      = {2019},
  booktitle = {Proc. SSW},
  pages     = {40--44},
}

@inproceedings{peng2018variational,
	title={Variational Discriminator Bottleneck: Improving Imitation Learning, Inverse {RL}, and {GAN}s by Constraining Information Flow},
	author={Xue Bin Peng and Angjoo Kanazawa and Sam Toyer and Pieter Abbeel and Sergey Levine},
	booktitle={Proc. ICLR},
	year={2019},
}

@article{cowen2019mapping,
  title={Mapping 24 emotions conveyed by brief human vocalization.},
  author={Cowen, Alan S and Elfenbein, Hillary Anger and Laukka, Petri and Keltner, Dacher},
  journal={American Psychologist},
  volume={74},
  number={6},
  pages={698},
  year={2019},
}

@article{SCHULLER20134,
title = {Paralinguistics in speech and language--{State}-of-the-art and the challenge},
journal = {Computer Speech \& Language},
volume = {27},
number = {1},
pages = {4-39},
year = {2013},
author = {Björn Schuller and Stefan Steidl and Anton Batliner and Felix Burkhardt and Laurence Devillers and Christian Müller and Shrikanth Narayanan},
}

@inproceedings{kanagawa25_ssw,
  title     = {{Multi-interaction TTS toward professional recording reproduction}},
  author    = {Hiroki Kanagawa and Kenichi Fujita and Aya Watanabe and Yusuke Ijima},
  year      = {2025},
  booktitle = {Proc. SSW},
  pages     = {110--116},
}

@ARTICLE{Schmidt23_showing,    
AUTHOR={Schmidt, Axel  and Deppermann, Arnulf },
TITLE={Showing and telling--{How} directors combine embodied demonstrations and verbal descriptions to instruct in theater rehearsals},
JOURNAL={Frontiers in Communication},          
VOLUME={7},
YEAR={2023},}

@inproceedings{fujita25_interspeech,
  title     = {Voice Impression Control in Zero-Shot {TTS}},
  author    = {Kenichi Fujita and Shota Horiguchi and Yusuke Ijima},
  year      = {2025},
  booktitle = {{Proc. Interspeech}},
  pages     = {4363--4367},
}

@inproceedings{ICLR2024_fed1ea8d,
 author = {Leng, Yichong and Guo, ZHifang and Shen, Kai and Ju, Zeqian and Tan, Xu and Liu, Eric and Liu, Yufei and Yang, Dongchao and zhang, leying and Song, Kaitao and He, Lei and Li, Xiangyang and zhao, sheng and Qin, Tao and Bian, Jiang},
 booktitle = {Proc. ICLR},
 pages = {57672--57688},
 title = {{PromptTTS 2}: {Describing} and Generating Voices with Text Prompt},
 year = {2024}
}

@inproceedings{ji-etal-2025-controlspeech,
    title = "{C}ontrol{S}peech: Towards Simultaneous and Independent Zero-shot Speaker Cloning and Zero-shot Language Style Control",
    author = "Ji, Shengpeng  and
      Chen, Qian  and
      Wang, Wen  and
      Zuo, Jialong  and
      Fang, Minghui  and
      Jiang, Ziyue  and
      Huang, Hai  and
      Wang, Zehan  and
      Cheng, Xize  and
      Zheng, Siqi  and
      Zhao, Zhou",
    booktitle = "Proc. ACL",
    year = "2025",
    pages = "6966--6981",
}

@inproceedings{Jin24_speechcraft,
author = {Jin, Zeyu and Jia, Jia and Wang, Qixin and Li, Kehan and Zhou, Shuoyi and Zhou, Songtao and Qin, Xiaoyu and Wu, Zhiyong},
title = {{SpeechCraft}: A Fine-Grained Expressive Speech Dataset with Natural Language Description},
year = {2024},
booktitle = {Proc. ACM MM},
pages = {1255–1264},
numpages = {10},
}

@INPROCEEDINGS{Yao24_promptvc,
  author={Yao, Jixun and Yang, Yuguang and Lei, Yi and Ning, Ziqian and Hu, Yanni and Pan, Yu and Yin, Jingjing and Zhou, Hongbin and Lu, Heng and Xie, Lei},
  booktitle={Proc. ICASSP}, 
  title={{PromptVC}: Flexible Stylistic Voice Conversion in Latent Space Driven by Natural Language Prompts}, 
  year={2024},
  pages={10571-10575},}

@misc{Ohmura25_libri,
      title={{LibriTTS-VI}: A Public Corpus and Novel Methods for Efficient Voice Impression Control}, 
      author={Junki Ohmura and Yuki Ito and Emiru Tsunoo and Toshiyuki Sekiya and Toshiyuki Kumakura},
      year={2025},
  howpublished= {arXiv},
  note={arXiv:2509.15626},
}

@inproceedings{
huang2024instructspeech,
title={{InstructSpeech}: Following Speech Editing Instructions via Large Language Models},
author={Rongjie Huang and Ruofan Hu and Yongqi Wang and Zehan Wang and Xize Cheng and Ziyue Jiang and Zhenhui Ye and Dongchao Yang and Luping Liu and Peng Gao and Zhou Zhao},
booktitle={Proc. ICML},
year={2024},
}

@inproceedings{
chen2026flexivoiceenablingflexiblestyle,
title={{FlexiVoice}: Enabling Flexible Style Control in Zero-Shot {TTS} with Natural Language Instructions},
author={Dekun Chen and Xueyao Zhang and Yuancheng Wang and Kenan Dai and Li Ma and Zhizheng Wu},
booktitle={Proc. ICLR},
year={2026},
}

@inproceedings{
liu2022flow,
title={Flow Straight and Fast: Learning to Generate and Transfer Data with Rectified Flow},
author={Xingchao Liu and Chengyue Gong and qiang liu},
booktitle={Proc. NeurIPS Workshop on Score-Based Methods},
year={2022},
}

@inproceedings{maini2024rephrasing,
title={Rephrasing the {Web}: A Recipe for Compute and Data-Efficient Language Modeling},
author={Pratyush Maini and Skyler Seto and He Bai and David Grangier and Yizhe Zhang and Navdeep Jaitly},
booktitle={Proc. ICLR Workshop},
year={2024},
}

@misc{gpt5_2,
    title={{Introducing GPT-5.2}}, 
    author={OpenAI},
    year={2025},
    note={Accessed:2026-02-09},
}

@inproceedings{langman25_interspeech,
  title     = {{HiFiTTS-2: A Large-Scale High Bandwidth Speech Dataset}},
  author    = {Ryan Langman and Xuesong Yang and Paarth Neekhara and Shehzeen Hussain and Edresson Casanova and Evelina Bakhturina and Jason Li},
  year      = {2025},
  booktitle = {Proc. {Interspeech}},
  pages     = {4778--4782},
}

@inproceedings{ma24d_interspeech,
  title     = {{WenetSpeech4TTS: A 12,800-hour Mandarin TTS Corpus for Large Speech Generation Model Benchmark}},
  author    = {Linhan Ma and Dake Guo and Kun Song and Yuepeng Jiang and Shuai Wang and Liumeng Xue and Weiming Xu and Huan Zhao and Binbin Zhang and Lei Xie},
  year      = {2024},
  booktitle = {Proc. {Interspeech}},
  pages     = {1840--1844},
}

@inproceedings{NEURIPS2023_voicebox,
title={Voicebox: Text-Guided Multilingual Universal Speech Generation at Scale},
 author = {Le, Matthew and Vyas, Apoorv and Shi, Bowen and Karrer, Brian and Sari, Leda and Moritz, Rashel and Williamson, Mary and Manohar, Vimal and Adi, Yossi and Mahadeokar, Jay and Hsu, Wei-Ning},
 booktitle = {Proc. NeurIPS},
 pages = {14005--14034},
 volume = {36},
 year = {2023}
}

@inproceedings{chen-etal-2025-f5,
    title = "F5-{TTS}: A Fairytaler that Fakes Fluent and Faithful Speech with Flow Matching",
    author = "Chen, Yushen  and
      Niu, Zhikang  and
      Ma, Ziyang  and
      Deng, Keqi  and
      Wang, Chunhui  and
      JianZhao, JianZhao  and
      Yu, Kai  and
      Chen, Xie",
    booktitle = "Proc. ACL",
    year = "2025",
    pages = "6255--6271",
}

@misc{hi-fi-captain,
  authour       =   {Takuma Okamoto and Yoshinori Shiga and Hisashi Kawai},
  title         =   {{Hi-Fi-CAPTAIN: High-fidelity and high-capacity conversational speech synthesis corpus developed by NICT}},
  url  =   {https://ast-astrec.nict.go.jp/en/release/hi-fi-captain/},
  year          =   {2023},
}

@inproceedings{baba2024utmosv2,
  title     = {The {T05} System for The {V}oice{MOS} {C}hallenge 2024: Transfer Learning from Deep Image Classifier to Naturalness {MOS} Prediction of High-Quality Synthetic Speech},
  author    = {Baba, Kaito and Nakata, Wataru and Saito, Yuki and Saruwatari, Hiroshi},
  booktitle = {Proc. SLT},
  year      = {2024},
  pages={818--824},
}

@inproceedings{fujita26_rie,
  title     = {Investigation for relative voice impression estimation},
  author    = {Kenichi Fujita and Yusuke Ijima},
  year      = {2026},
  booktitle = {Proc. {Speech Prosody}},
  pages     = {486--490},
}

@INPROCEEDINGS{Yong26_ov,
  author={Ren, Yong and Yi, Jiangyan and Tao, Jianhua and Sun, Haiyang and Wen, Zhengqi and Gu, Hao and Xu, Le and Bai, Ye},
  booktitle={Proc. ICASSP}, 
  title={{OV-InstructTTS}: Towards Open-Vocabulary Instruct Text-to-Speech}, 
  year={2026},
  pages={16482-16486},}

@INPROCEEDINGS{Yun26_isse,
  author={Chen, Yun and Chen, Qi and Dai, Zheqi and Singh, Arshdeep and Jackson, Philip J.B. and Plumbley, Mark D.},
  booktitle={Proc. ICASSP}, 
  title={{ISSE}: An Instruction-Guided Speech Style Editing Dataset and Benchmark}, 
  year={2026},
  volume={},
  number={},
  pages={19482-19486},
}

@inproceedings{
chen2024mllmasajudge,
title={{MLLM}-as-a-Judge: Assessing Multimodal {LLM}-as-a-Judge with Vision-Language Benchmark},
author={Dongping Chen and Ruoxi Chen and Shilin Zhang and Yaochen Wang and Yinuo Liu and Huichi Zhou and Qihui Zhang and Yao Wan and Pan Zhou and Lichao Sun},
booktitle={Proc. ICML},
year={2024},
}

@inproceedings{li-etal-2025-generation,
    title = "From Generation to Judgment: Opportunities and Challenges of {LLM}-as-a-judge",
    author = "Li, Dawei  and
      Jiang, Bohan  and
      Huang, Liangjie  and
      Beigi, Alimohammad  and
      Zhao, Chengshuai  and
      Tan, Zhen  and
      Bhattacharjee, Amrita  and
      Jiang, Yuxuan  and
      Chen, Canyu  and
      Wu, Tianhao  and
      Shu, Kai  and
      Cheng, Lu  and
      Liu, Huan",
    booktitle = "Proc. EMNLP",
    year = "2025",
    pages = "2757--2791",
}

@misc{speechbrain,
  title={{SpeechBrain}: A General-Purpose Speech Toolkit},
  author={Mirco Ravanelli and Titouan Parcollet and Peter Plantinga and Aku Rouhe and Samuele Cornell and Loren Lugosch and Cem Subakan and Nauman Dawalatabad and Abdelwahab Heba and Jianyuan Zhong and Ju-Chieh Chou and Sung-Lin Yeh and Szu-Wei Fu and Chien-Feng Liao and Elena Rastorgueva and François Grondin and William Aris and Hwidong Na and Yan Gao and Renato De Mori and Yoshua Bengio},
  year={2021},
  howpublished={arXiv},
  note={arXiv:2106.04624}
}

@misc{
    modernbert-ja,
    author = {Tsukagoshi, Hayato and Li, Shengzhe and Fukuchi, Akihiko and Shibata, Tomohide},
    title = {{ModernBERT-Ja}},
    url = {https://huggingface.co/collections/sbintuitions/modernbert-ja-67b68fe891132877cf67aa0a},
    year = {2025},
}

@article{zhou2021emotional,
title = {Emotional voice conversion: Theory, databases and {ESD}},
journal = {Speech Communication},
volume = {137},
pages = {1-18},
year = {2022},
author = {Kun Zhou and Berrak Sisman and Rui Liu and Haizhou Li},
}

@ARTICLE{yamagishi2005HMM,
  author={Junichi Yamagishi and Koji Onishi and Takashi Masuko and Takao Kobayashi},
  journal={IEICE TRANSACTIONS on Information}, 
  title={Acoustic Modeling of Speaking Styles and Emotional Expressions in {HMM}-Based Speech Synthesis}, 
  year={2005},
  volume={E88-D},
  number={3},
  pages={502-509},
}

\end{document}